\documentclass[preprint,superscriptaddress]{revtex4}

\usepackage{graphicx}
\usepackage{color}
\usepackage{url}
\usepackage{gensymb}
\usepackage{epsf}

\begin{document}

\title{Collapse of the Fermi surface and fingerprints of order in the pseudogap state of cuprate superconductors} 

\author{Takeshi~Kondo}
\affiliation{Ames Laboratory and Department of Physics and Astronomy, Iowa State
University, Ames, IA 50011, USA}
\author{Ari~D.~Palczewski}
\affiliation{Ames Laboratory and Department of Physics and Astronomy, Iowa State
University, Ames, IA 50011, USA}
\author{Yoichiro~Hamaya}
\affiliation{Department of Crystalline Materials Science, Nagoya University, Nagoya
464-8603, Japan}
\author{Tsunehiro~Takeuchi}
\affiliation{Department of Crystalline Materials Science, Nagoya University, Nagoya
464-8603, Japan}
\affiliation{EcoTopia Science Institute, Nagoya University, Nagoya 464-8603, Japan}
\author{J.~S.~Wen}
\affiliation{Condensed Matter Physics and Materials Science Department, Brookhaven
National Laboratory, Upton, New York 11973, USA }
\author{Z.~J.~Xu}
\affiliation{Condensed Matter Physics and Materials Science Department, Brookhaven
National Laboratory, Upton, New York 11973, USA }
\author{Genda~Gu}
\affiliation{Condensed Matter Physics and Materials Science Department, Brookhaven
National Laboratory, Upton, New York 11973, USA}
\author{Adam~Kaminski}
\affiliation{Ames Laboratory and Department of Physics and Astronomy, Iowa State
University, Ames, IA 50011, USA}

\maketitle

{\bf The Fermi surface in the pseudogap\cite{Warren,Takigawa,Homes,HongPseudogap,LoeserPseudogap} state of cuprates is highly unusual because it appears to consist of disconnected segments called arcs\cite{NormanNature}. Their very existence challenges the traditional concept of a Fermi surface as closed contours of gapless excitations in momentum space. The length of the arcs in the pseudogap state was thought to linearly increase with temperature, pointing to the existence of a nodal liquid state below $T^*$\cite{KanigelNP,Chatterjee}. These results were interpreted as an interplay of a d-wave pairing gap\cite{VallaScience} and strong scattering\cite{ChubukovScattering}. Understanding the properties of the arcs is a pre-requisite to understanding the origin of the pseudogap\cite{MikeFoe, Emery, Chatterjee, DavisFingerprint, Johnson, ShenScience, Raman, KondoTwogap, EricTwogap, Rustem, KondoCompetition} and the physics of the cuprates. Here we  use a novel approach to  detect the energy gaps based on the temperature dependence of the density of states. With a significantly improved sensitivity, we demonstrate that the arcs form rapidly at $T^*$ and their length remains surprisingly constant over an extended temperature range between \boldmath{$T^{*}$} and \boldmath{$T_{\rm arc}$}, consistent with the presence of an ordered state below \boldmath{$T^{*}$}. These arcs span fixed points in the momentum space defining a set of wave vectors, which are the fingerprints of the ordered state that causes the pseudogap.}

Traditional methods for determining the presence of an energy gap in ARPES spectra \cite{HongPseudogap,LoeserPseudogap,NormanNature,KanigelNP} rely on a lineshape analysis such as a shift in the leading edge or a dip in symmetrized spectra\cite{NormanNature}. In the cuprates however, these features are poorly defined (especially above $T_{\rm c}$) because 
the spectral peaks are very broad in optimally and underdoped samples. This makes the detection of very small energy gaps and partial gaps (such as the pseudogap, which affects only part of the spectral weight) difficult or impossible to do. Even gapped ARPES spectra may appear to have a single peak at $E_{\rm F}$ after symmetrization if they are sufficiently broad. One of the most sensitive ways to detect of the opening of an energy gap is via the density of states (DOS($E_{\rm F}$)) at the Fermi energy. In a gapless state, this quantity is independent of temperature, because unlike the spectral function it does not depend on the electron lifetime.  The opening of an energy gap at the Fermi energy leads to a decrease of DOS($E_{\rm F}$). Studying this quantity as a function of temperature therefore provides a very sensitive measure of a gap opening, which is quantitative and objective. To obtain momentum dependent information we use the area of momentum distribution curve (MDC) at the Fermi energy (D$_{\rm MDC}$($E_{\rm F}$)) which represents a contribution to the DOS($E_{\rm F}$) from a small slice of the Brillouin zone. This new approach requires very high quality data and temperature stability, but at the same time it allows us to obtain momentum resolved information about the opening of an energy gap with significantly higher accuracy than previously possible. Details of the sample preparation and ARPES measurements are provided in the Supplementary Information.

In Fig. 1 we demonstrate the above procedure using data at the node (gapless) and away from the node (gapped) from a Bi2201 sample and compare it to the traditional method of symmetrization. Fig. 1a shows the MDCs along the nodal cut (see inset of Fig. 1b) measured at various temperatures from below the superconducting transition temperature ($T_{\rm c}$) to above the $T^*$. The MDC peak, which is observed at $k_{\rm F}$, broadens with increased temperature, as the electron lifetime shortens due to scattering. The area of the MDCs, D$_{\rm MDC}$($E_{\rm F}$), remains constant with temperature as expected for a partial contribution to the DOS in the absence of an energy gap (Fig. 1b). This behavior changes dramatically away from the node, where an energy gap opens.  Fig. 1c shows the MDCs along a momentum cut close to the antinode (see inset of Fig.1d). At high temperatures above $T^*$,  the peak  width increases, while the area remains constant.  Below $T^*$, the area of the peaks starts to decrease, indicating the opening of an energy gap. This is even more evident, when examining the D$_{\rm MDC}$($E_{\rm F}$)  as a function of temperature shown in Fig. 1d. This quantity is constant at high temperatures, then decreases below $T^*$.  The reduction of D$_{\rm MDC}$($E_{\rm F}$) below $T^*$ signifies the opening of an energy gap - in this case the pseudogap (indicated by a blue arrow). We note that there are number of features in D$_{\rm MDC}$ below the temperature at which the gap opens. They very likely contain important information related to the paring \cite{KondoPairing} and the superfluid density. They are, however, beyond the scope of this work and will be analyzed carefully in the future.

For a comparison, we plot the symmetrized energy distribution curves (EDCs) at $k_{\rm F}$ for several temperatures in Fig.1e. This is the traditional way to detect the energy gap\cite{NormanNature,KanigelNP}.  The single peak present at high temperatures develops a dip on cooling, which was previously interpreted as a signature of a gap opening. It is clear that the temperature ($T_{\rm peak}$) is considerably lower than the $T^*$, because smaller gaps even if present do not always produce a dip in the symmetrized spectra. The traditional approach fails to detect small energy gaps and lacks the sensitivity necessary to reveal the real nature of the Fermi arcs. Careful investigation of D$_{\rm MDC}$($E_{\rm F}$,$T$) is essential to achieve this goal.

In order to study the temperature evolution of the Fermi arc, we carefully measured D$_{\rm MDC}$($E_{\rm F}$,$T$) for a number of Fermi momentum points.
The sample orientation was adjusted in each case, so the cuts were normal to the Fermi surface (see insets of Fig. 2).
The left and right panels in Fig.  2 show the data obtained from optimally doped Bi2201 ($T_{\rm c}$=32K) and Bi2212 ($T_{\rm c}$=93K), respectively (raw MDC data is included in the Supplementary Information).  The data in the top panels was obtained at the antinode, while the lower panels show data towards the node. The D$_{\rm MDC}$($E_{\rm F}$,$T$) starts to decrease upon cooling at $T^*$ in the top four panels for $0^\circ \leq\phi\leq19^\circ $, indicated by blue arrow. For the remaining Fermi momentum points closer to the node, the D$_{\rm MDC}$($E_{\rm F}$,$T$) remains constant down to a much lower temperature. This signifies the simultaneous opening of the pseudogap  for a range of Fermi momentum points up to $\phi\leq19^\circ $ at $T^*$. The Fermi surface therefore collapses very abruptly into arcs just below  $T^*$. 
 The lack of a variation in D$_{\rm MDC}$($E_{\rm F}$,$T$) for $\phi>19^\circ $ down to $T_{\rm arc}$ demonstrates that the length of the arcs remains constant in this temperature range. On further cooling below $T_{\rm arc}$, D$_{\rm MDC}$($E_{\rm F}$,$T$) starts to simultaneously decrease over the remaining segments of the FS. This signifies the second collapse of the remaining segments of the Fermi surface (arcs) below $T_{\rm arc}$ into a very small arc or most likely the point node of a $d$-wave paired state. We summarize these results in Fig. 3 by plotting the gap opening temperatures as a function of the Fermi surface angle and the Fermi arc length as a function of temperature, based on the data in Fig. 2.

Figure 4 shows a schematic diagram demonstrating the process of forming the arcs and their subsequent collapse. Our most important finding is the double collapse of the Fermi surface on cooling first at $T^*$ and then at $T_{\rm arc}$. This result clearly contradicts the nodal liquid scenario \cite{KanigelNP,Chatterjee}, where the Fermi arcs expand linearly with temperature. We emphasize that the traditional lineshape analysis underestimates the temperature at which the pseudogap opens. To confirm this, we repeated a traditional symmetrization analysis for our data from Fig.2, and demonstrated an apparent linear expansion of the arcs with temperature (see Supplementary Information). The remaining open issue is whether a tiny Fermi arc or a single point exist very close to the node below $T_{\rm pair}$. At this time, due to the limits of the data accuracy we can only impose limits on the length of this arc to about  20\% of the normal state Fermi surface.

The Fermi momentum points at the tip of the pseudogap arcs coincide with the ends of the parallel segments of the FS rather than at the antiferromagnetic zone boundary (AFZB). The vectors connecting these points such as (0, 0.4), (1.4,0), (1.4, 0.4) are likely the key to determining the nature of the ordered state underlying the pseudogap phenomenon in cuprates. Earlier STM studies \cite{EricCDW,EricFS} proposed a scenario for the pseudogap based on a charge density wave with a (0, 0.4) ordering vector. The strong energy dependence of the checkerboard pattern \cite{Hanaguri} in STM is however inconsistent with the signatures of a classical charge density wave. The other two vectors however may be related to a nematic state reported by separate STM study\cite{DavisNematic}. So far we have not been able to identify a theory consistent with the above set of vectors, however we hope that this information will lead to the development of a correct theory of the pseudogap state.

\textbf{Corresponding Author}\newline
Correspondence to: A. Kaminski. e-mail: kaminski@ameslab.gov or T. Kondo. e-mail: kondo@ameslab.gov.

\textbf{Acknowledgements}\newline
We thank J\"org~Schmalian Schmalian, Mike Norman and Andrey V. Chubukov for useful discussions. 
This work was supported by Basic Energy Sciences, US DOE. The Ames Laboratory is
operated for the US DOE by Iowa State University under Contract No. DE-AC02-07CH11358. Work at Brookhaven is supported by the US DOE under Contract No. DE-AC02-98CH10886. JSW and ZJX are supported by the Center for Emergent Superconductivity, an Energy Frontier Research Center funded by the US DOE, Office of Science.

{\bf Author Contributions} \newline
 T.K. and A.K. designed the experiment. T.K., Y.H., T.T., J.S.W, G.Z.J.X, and G.G grew the high-quality single crystals. T.K. and A.D.P acquired the experimental data and T.K. performed the data analysis. T.K. and A.K. wrote the manuscript. All authors discussed the results and commented on the manuscript. 
 
\textbf{Additional information}\newline
The authors declare no competing financial interests. Reprints and permissions
information is available online at http://npg.nature.com/reprintsandpermissions.
Correspondence and requests for materials should be addressed to A. K. or T. K.

\clearpage

\begin{figure} 
\includegraphics[width=6in]{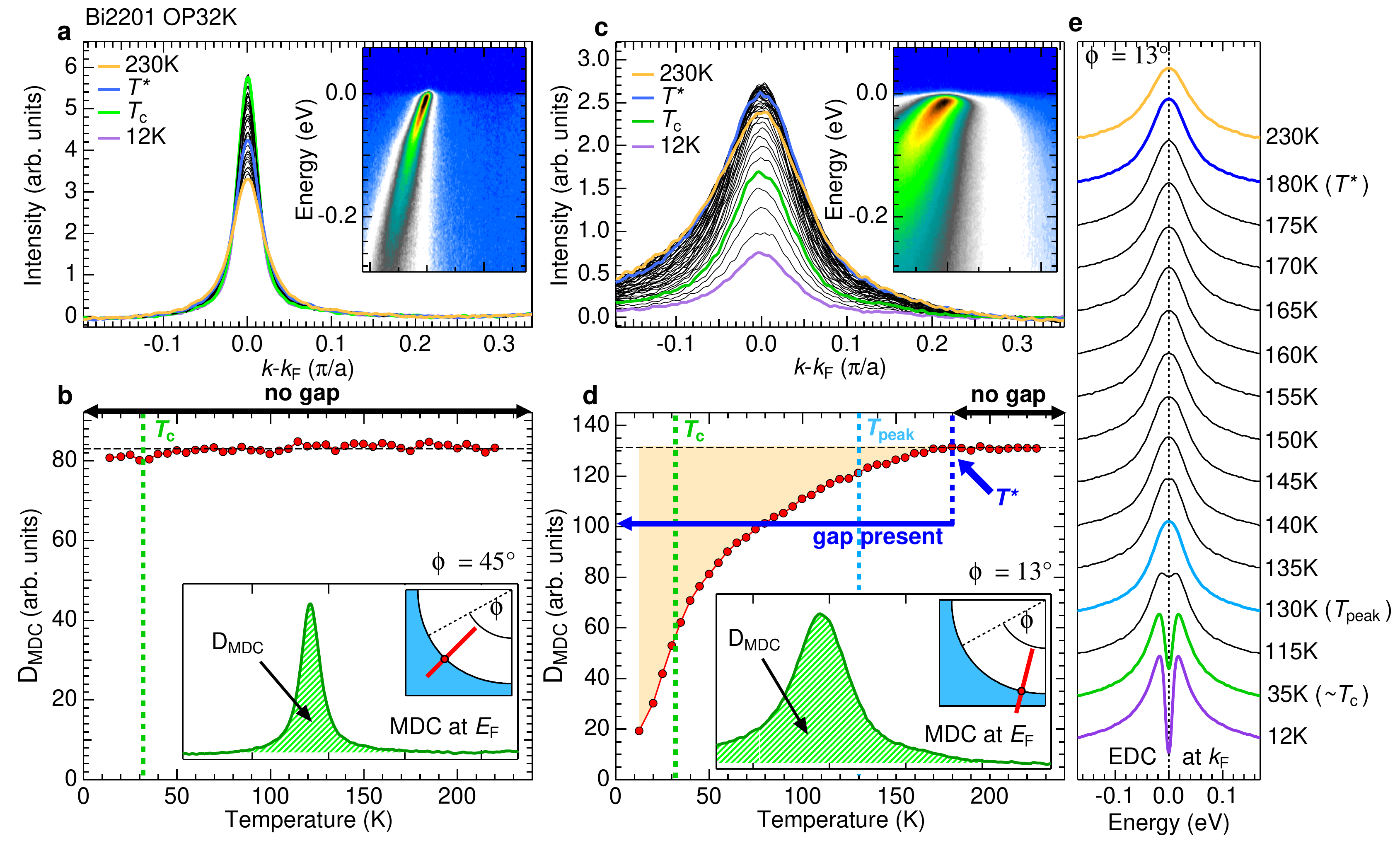}
\caption{ Determination of the partial density of state at $E_{\rm F}$, D$_{\rm MDC}$($E_{\rm F}$), from the ARPES data of optimally doped Bi2201 ($T_{\rm c}$=32K). 
{\bf a,} MDCs measured along the nodal direction over a wide range of temperatures. Inset shows a dispersion image at the lowest temperature (12K). 
{\bf b,} The temperature dependence of D$_{\rm MDC}$($E_{\rm F}$) estimated from the area of MDCs in a. The observed momentum cut and estimated area of MDC are 
demonstrated in the inset. 
{\bf c,} Same data as in a, but measured along a momentum cut slightly off the antinode. Inset shows a dispersion image at the lowest temperature (12K).  
{\bf d,} The temperature dependence of D$_{\rm MDC}$($E_{\rm F}$) estimated from the area of MDCs in c. The observed momentum cut and estimated area of MDC are shown in the inset. The pseudogap temperature ($T^*$) is defined as the temperature where the D$_{\rm MDC}$($E_{\rm F}$,$T$) starts to decrease on cooling.
{\bf e,} Symmetrized EDCs at $k_{\rm F}$ for several temperatures, obtained from the data of c. The $T_{\rm peak}$ is defined as the temperature where two peaks in the spectrum merge to one peak at elevated temperatures. }
\label{fig1}
\end{figure}

\newpage

\begin{figure} 
\includegraphics[width=2.5in]{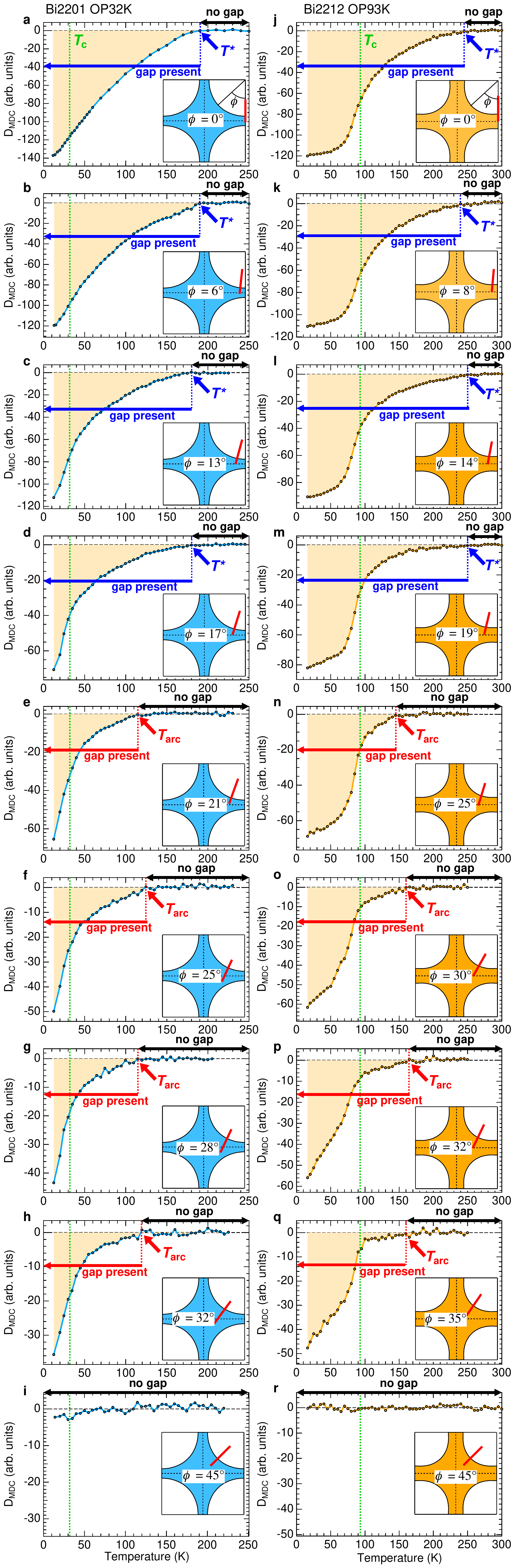}
\caption{ }
\label{fig2}
\end{figure}

\newpage
\clearpage
FIG. 2. Angle dependence of the partial density of state, D$_{\rm MDC}$($E_{\rm F}$).
{\bf a--i,} for optimally doped Bi2201 with $T_{\rm c}$=32K.
{\bf j--r} for optimally doped Bi2212 with $T_{\rm c}$=93K. The top panels correspond to the antinodal direction. Lower panels show data along cuts approaching the nodal direction. The inset of each panel shows the FS and the location of the cut, normal to the FS. The pseudogap temperature ($T^*$) and the temperature above which the arc exists ($T_{\rm arc}$) are indicated by blue and red arrows, respectively.

\newpage

\begin{figure} 
\includegraphics[width=6in]{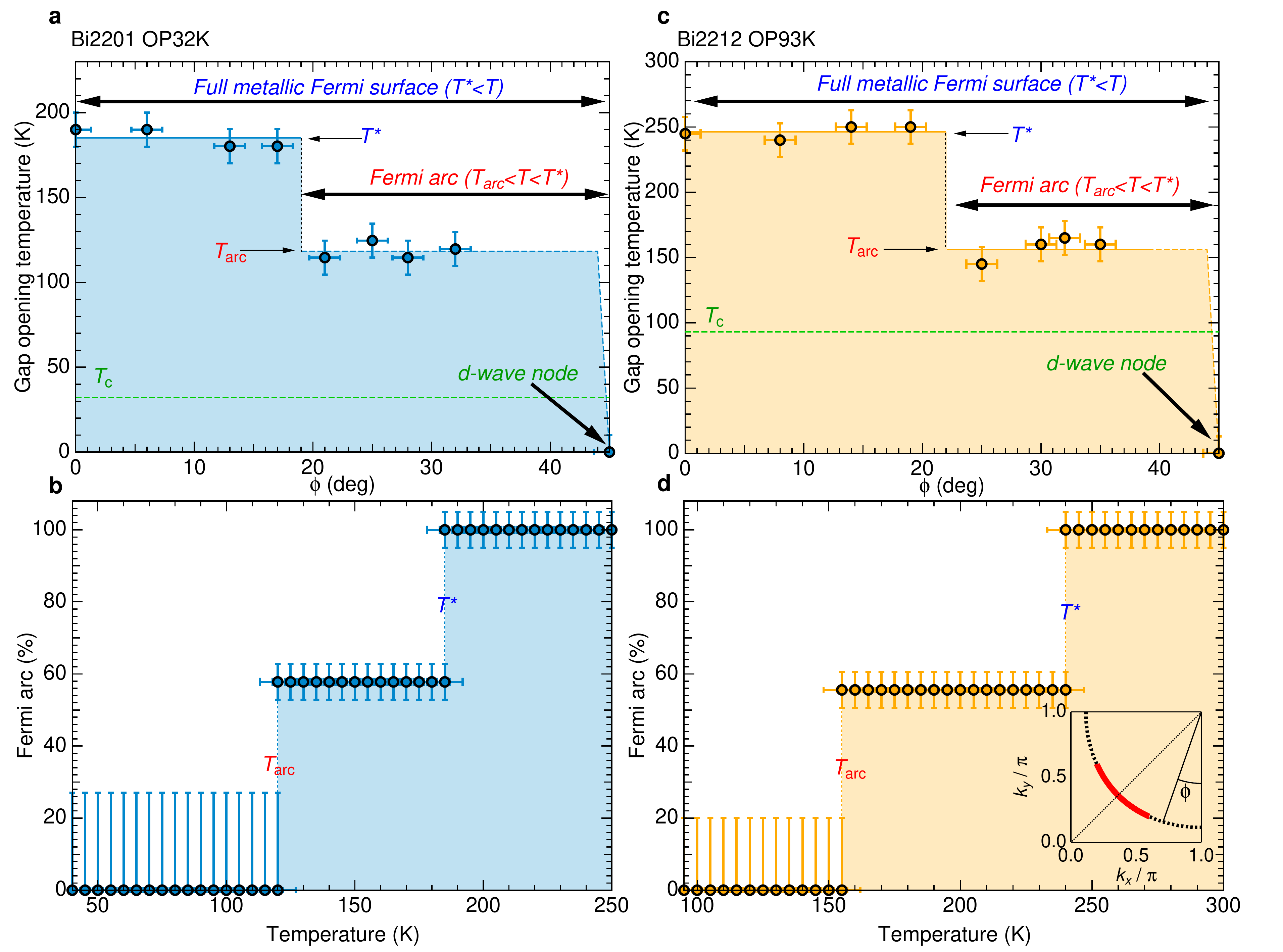}
\caption{
Momentum dependence of the temperatures at which gap opens and the length of the Fermi arc determined from ARPES data shown in Fig. 2.
{\bf a,} Gap opening temperatures as a function of Fermi surface angle for optimally doped Bi2201.
{\bf b,} Temperature dependence of the arc length. Abrupt changes in the length of the arcs occur at  $T_{\rm arc}$ and $T^*$. Their length remains constant for $T_{\rm arc}  < T < T^* $.
{\bf c--d,} same as a and b, but for optimally doped Bi2212.
}
\label{fig3}
\end{figure}

\newpage

\begin{figure} 
\includegraphics[width=6in]{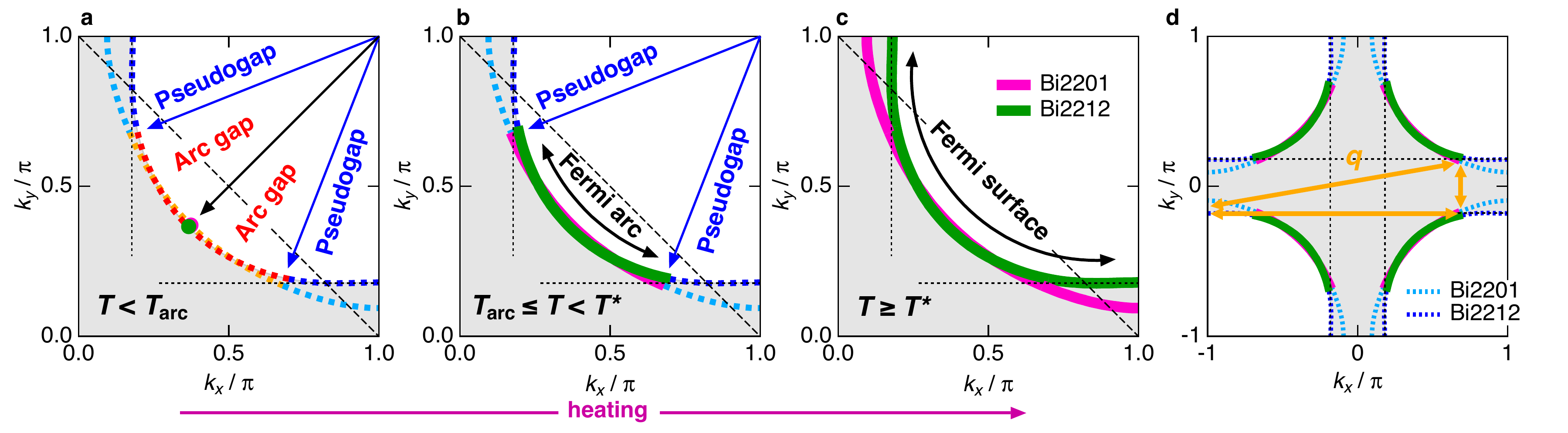}
\caption{
Schematic diagram of the temperature evolution of the Fermi surface in cuprates revealed by the new method. 
{\bf a--c,} The Fermi surface for three key temperature ranges. The antiferromagnetic zone boundary (AFZB) is marked with a black dashed line. The black dotted lines are the guide to eyes for parallel segments of Fermi surface. {\bf d,} The Fermi surface for the Bi2212 and Bi2201 determined from our ARPES data.  The key \boldmath{$q$}  vectors connecting the nested sections of the FS are indicated with arrows. The black dotted lines a the guide to the eye for parallel segments of Fermi surface.
}\label{fig4}
\end{figure}

\end{document}


\title{On-line supplementary information for

Collapse of the Fermi surface and fingerprints of order in the pseudogap state of cuprate superconductors}

\author{Takeshi~Kondo}
\affiliation{Ames Laboratory and Department of Physics and Astronomy, Iowa State
University, Ames, IA 50011, USA}

\author{Ari~Palczewski}
\affiliation{Ames Laboratory and Department of Physics and Astronomy, Iowa State
University, Ames, IA 50011, USA}

\author{Yoichiro~Hamaya}
\affiliation{Department of Crystalline Materials Science, Nagoya University, Nagoya
464-8603, Japan}

\author{K.~Ogawa}
\affiliation{Department of Crystalline Materials Science, Nagoya University, Nagoya
464-8603, Japan}

\author{Tsunehiro~Takeuchi}
\affiliation{Department of Crystalline Materials Science, Nagoya University, Nagoya
464-8603, Japan}
\affiliation{EcoTopia Science Institute, Nagoya University, Nagoya 464-8603, Japan}

\author{J.~S.~Wen}
\affiliation{Condensed Matter Physics and Materials Science Department, 
Brookhaven National Laboratory, Upton, New York 11973, USA}

\author{G.~Z.~J.~Xu}
\affiliation{Condensed Matter Physics and Materials Science Department, 
Brookhaven National Laboratory, Upton, New York 11973, USA}

\author{Genda~Gu}
\affiliation{Condensed Matter Physics and Materials Science Department, 
Brookhaven National Laboratory, Upton, New York 11973, USA}

\author{Adam~Kaminski}
\affiliation{Ames Laboratory and Department of Physics and Astronomy, Iowa State
University, Ames, IA 50011, USA}
\date{\today }

\date{\today}

\maketitle

\section {Samples and Experimental method of ARPES}
Optimally doped Bi$_2$Sr$_2$CaCu$_2$O$_{8+\delta}$ (Bi2212) single crystals with $T_{\rm c}$=93K (OP93K) and \\
(Bi,Pb)$_2$(Sr,La)$_2$CuO$_{6+\delta}$ (Bi2201) single crystals with $T_{\rm c}$=32K (OP32K) were 
grown by the conventional floating-zone (FZ) technique.  
We partially substituted Pb for Bi in Bi2201 to suppress the modulation in the BiO plane, 
and avoid contamination of the ARPES signal with diffraction images due to the superlattice: the outgoing photoelectrons are diffracted at the modulated BiO layer, 
creating multiple images of the bands and Fermi surface that are shifted in momentum. The modulation-free samples enable us to precisely analyze the ARPES spectra. 

ARPES data was acquired using a laboratory-based system consisting of a Scienta SES2002 electron analyzer and GammaData Helium UV lamp. All data were acquired using the HeI line with a photon energy of 21.2 eV. The angular resolution was $0.13^\circ$ and $\sim 0.5^\circ$ along and perpendicular to the direction of the analyzer slits, respectively.  
The energy corresponding to the chemical potential was determined from the Fermi edge of a polycrystalline Au reference in electrical contact with the sample. The energy resolution was set at $\sim10$meV - confirmed by measuring the energy width between 90$\%$ and 10 $\%$ of the Fermi edge from the same Au reference. 
Custom designed refocusing optics enabled us to accumulate  high statistics spectra in a short period of time with no sample surface aging from the absorption or loss of oxygen. Special care was taken to purify the helium gas supply for the UV source to remove even the smallest trace of contaminants that could contribute to surface contamination. Typically no changes in the spectral lineshape of samples were observed in consecutive measurements performed over several days. We constructed a sample manipulator with the tilt and azimuth motions mounted on a two stage closed cycle He refrigerator. To measure the partial density of states along a direction perpendicular to the FS we controlled the sample orientation in-situ. Measurements were performed on several samples and we confirmed that all yielded consistent results. 

\section {Normalization of ARPES spectra}
To reduce the noise level when collecting data over an extended temperature range, we normalized the band dispersion map (see Fig. S1 e) at each temperature, to the whole area of the symmetrized energy distribution curve (EDC) at the $k_{\rm F}$ point in the range of -0.4eV $\leq E\leq $ 0.4eV.  Here we demonstrate that a particular normalization scheme does not affect the results.
Fig. S1 a and c show the symmetrized EDCs at the $k_{\rm F}$ point near the antinode measured at various temperatures.
The spectra in panel a are normalized to the whole area of each spectrum in the range of  -0.4eV to 0.4eV, while spectra in panel c are normalized to the intensity within $\pm$50 meV centered at -0.35eV.
Please note that in the data normalized over the wide energy range (panel a), the intensity at higher energies (e. g. -0.35eV) is almost constant for all temperatures. 
It is, therefore, expected that the results obtained by the two different normalization schemes are consistent.
The data normalized using these two procedures at each temperature was used to extract MDC area at $E_{\rm F}$ (i. e. D$_{\rm MDC}$($E_{\rm F}$) vs $T$)  and is shown in Fig. 1S f.
As expected, the temperature dependence of this quantity is very similar for the two normalization schemes. The two curves are almost identical with the same onset of the reduction in D$_{\rm MDC}$($E_{\rm F}$) on cooling at  $T^*$=180K (the pseudogap temperature). 
This allows us to conclude that the results we present, including the determination of the gap opening temperatures, do not dependent on the choice of a normalization scheme.

The intensity of spectral function is expected to be zero far from the location of the band. However, the ARPES intensity at $E_{\rm F}$ in such a momentum region is usually finite. This background signal is due to  photoelectrons that have lost their momentum information during the photoemission process via  elastic scattering\cite{AdamBackgr}.  We subtract this intensity from the experimental momentum distribution curves (MDCs) before estimating the MDC area for each temperature.  We have checked the validity of this treatment by performing the same analysis for many data sets from different samples. The intensity due to the elastic scattering varies slightly from sample to sample because elastic scattering is very sensitive to the condition of the sample surface. The behavior of D$_{\rm MDC}$($E_{\rm F}$) (area of MDC) estimated before and after such a procedure is nevertheless consistent for all samples. The checks described above validate all steps of our analysis of the ARPES data. 
 
 \section {Comparison of the results from quantitative analysis with lineshape analysis}
In Fig. S2 and S3 we use our high--quality data to demonstrate the inadequacy of the traditional lineshape analysis whereby the spectral peak in the symmetrized data is tracked. This method results in an underestimation of the temperature at which the gap closes. 
 Figure S2 and S3 show the symmetrized  energy distribution curves (EDCs) measured at various temperatures from 
 Bi2201 and Bi2212 samples, respectively. From left to right, we plot the data from the antinode to the node. 
 At elevated temperatures, the two peaks in the spectra merge into one (a red curve) at a characteristic temperature ($T_{\rm peak}$). 
It is easily seen that the $T_{\rm peak}$ monotonically decreases toward the node. 
 As a result, the portion of the Fermi surface where the  spectra have a peak at $E_{\rm F}$ emerges centered at the node, and it 
  expands toward the antinode with increasing temperature.
  In the inset of Fig. S2 and S3 we estimate the length of these "artificial" Fermi arc as a function of temperature. 
Such plotted arc lengths are similar to an earlier report \cite{KanigelNP}. The length of arcs increases almost linearly with temperature above $T_{\rm c}$ up to $T^*$.
We emphasize that at momentum points, where the traditional lineshape method  determines the gap to be closed, the partial density of states is still increasing with temperature. This signifies that the energy gap is still present, but is not detected using symmetrization and is most likely due to thermal broadening or a rapid increase of scattering above $T_{\rm c}$. 
The biggest advantage of utilizing the partial density of states at $E_{\rm F}$, D$_{\rm MDC}$($E_{\rm F},T$),  is that it is not affected by the thermal broadening and is very sensitive to presence of even a very small energy gap. From the reduction in the D$_{\rm MDC}$($E_{\rm F},T$), one can estimate the gap opening temperature
much more precisely than with the traditional approach.

 \section { Raw data used in analysis shown in Fig. 2.}
 In Fig. S4 a-i and S5 a-i, we present the MDCs along many directions of Bi2201 OP32K and Bi2212 OP93K, respectively, used to estimate the partial density of states, D$_{\rm MDC}$($E_{\rm F}$),  (shown in Fig.2 of the main text). In the inset of each panel we plot the band dispersion image for the lowest temperature (12K). The panels j-r of Fig. S4  and S5 show the same data as in Fig. 2. In each panel,  the gap closing temperature estimated from the spectral peak position ($T_{\rm peak}$) and the density of state ($T_{\rm DOS}$) are indicated.

\clearpage
\newpage

\renewcommand{\thefigure}{S\arabic{figure}}
\begin{figure*}
\includegraphics[width=6in]{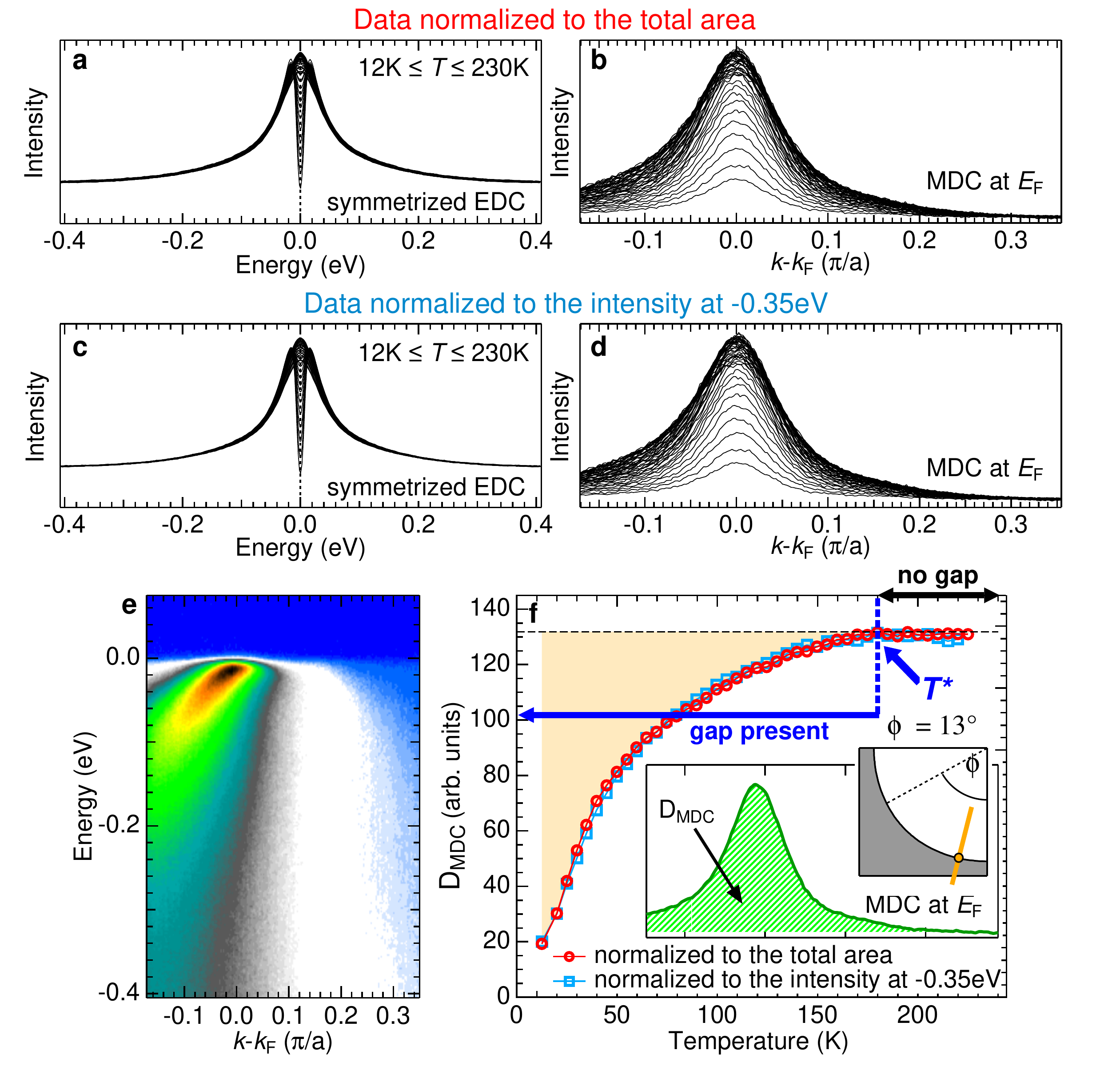}
\label{fig1}
\end{figure*}

\noindent {\bf Fig. S1.} 
Verification of normalization procedure.  
{\bf a}, Symmetrized EDCs at the $k_F$ close to the antinode [see inset of {\bf f}] measured over a wide temperature range from below $T_{\rm c}$ to above $T^*$. All spectra are normalized to the total area integrated from -0.4eV to 0.4eV. 
{\bf b}, MDCs at various temperatures along the momentum cut indicated in the inset of {\bf f}, extracted from the band dispersion maps normalized to 
the total area of symmetrized EDCs at $k_F$ for each temperature. 
{\bf c}, Same spectra as in {\bf a} normalized to the intensity within a narrow energy window ($\pm$50meV) centered at -0.35eV.
{\bf d}, MDCs at various temperatures, extracted from the dispersion map normalized to the intensity of spectrum at $k_F$ within a narrow energy window ($\pm$50meV) centered at -0.35eV.  
{\bf e}, Band dispersion map at the lowest temperature (12K).
{\bf f}, Two sets of D$_{\rm MDC}$($E_{\rm F}$) vs $T$ curves [area of MDC], obtained using two different normalization schemes. The inset shows 
the estimated area of MDC and the Fermi surface with the momentum cut measured.

\clearpage
\newpage

\renewcommand{\thefigure}{S\arabic{figure}}
\begin{figure*}
\includegraphics[width=6.5in]{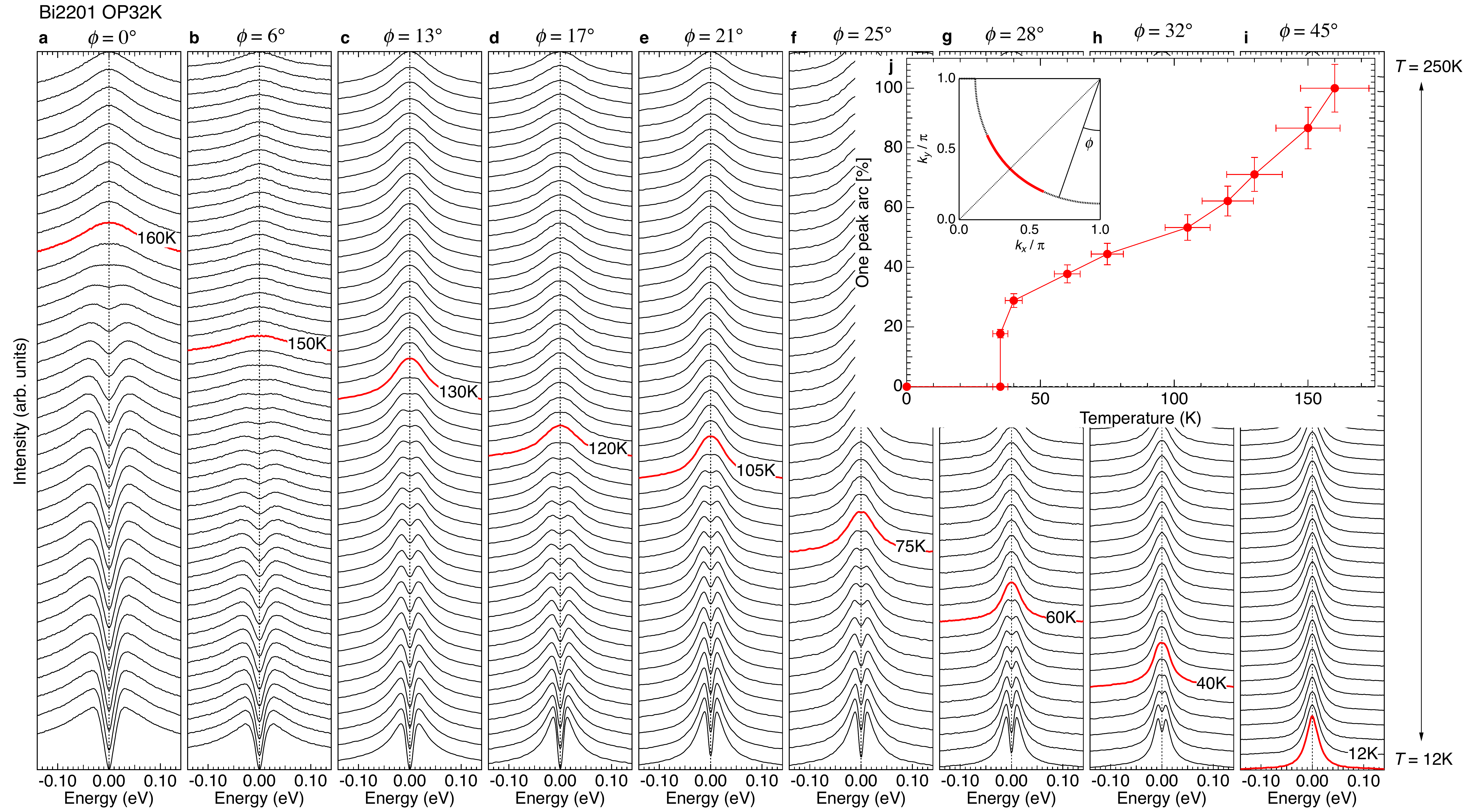}
\label{fig2}
\end{figure*}

\noindent {\bf Fig. S2.} 
The "artificial" Fermi arc of Bi2201 OP 32K obtained by tracking spectral peak. 
In each panel, the symmetrized EDCs measured  over a wide temperature range from below $T_{\rm c}$ to above $T^*$ are plotted.
From the left ({\bf a}) to the right ({\bf i}) panel, the measured $k_F$ point changes from the antinode to the node. 
The Fermi angle of the measured $k_F$ point [$\phi$: see inset of {\bf j}] is described on the top of each panel. 
The spectrum where the two peaks  merge to one peak at elevated temperatures is indicated with a red color. 
{\bf j}, The length of Fermi surface ("artificial" Fermi arc) where one peak structure is observed is plotted as a function of temperature.

\clearpage
\newpage

\begin{figure*}
\includegraphics[width=6.5in]{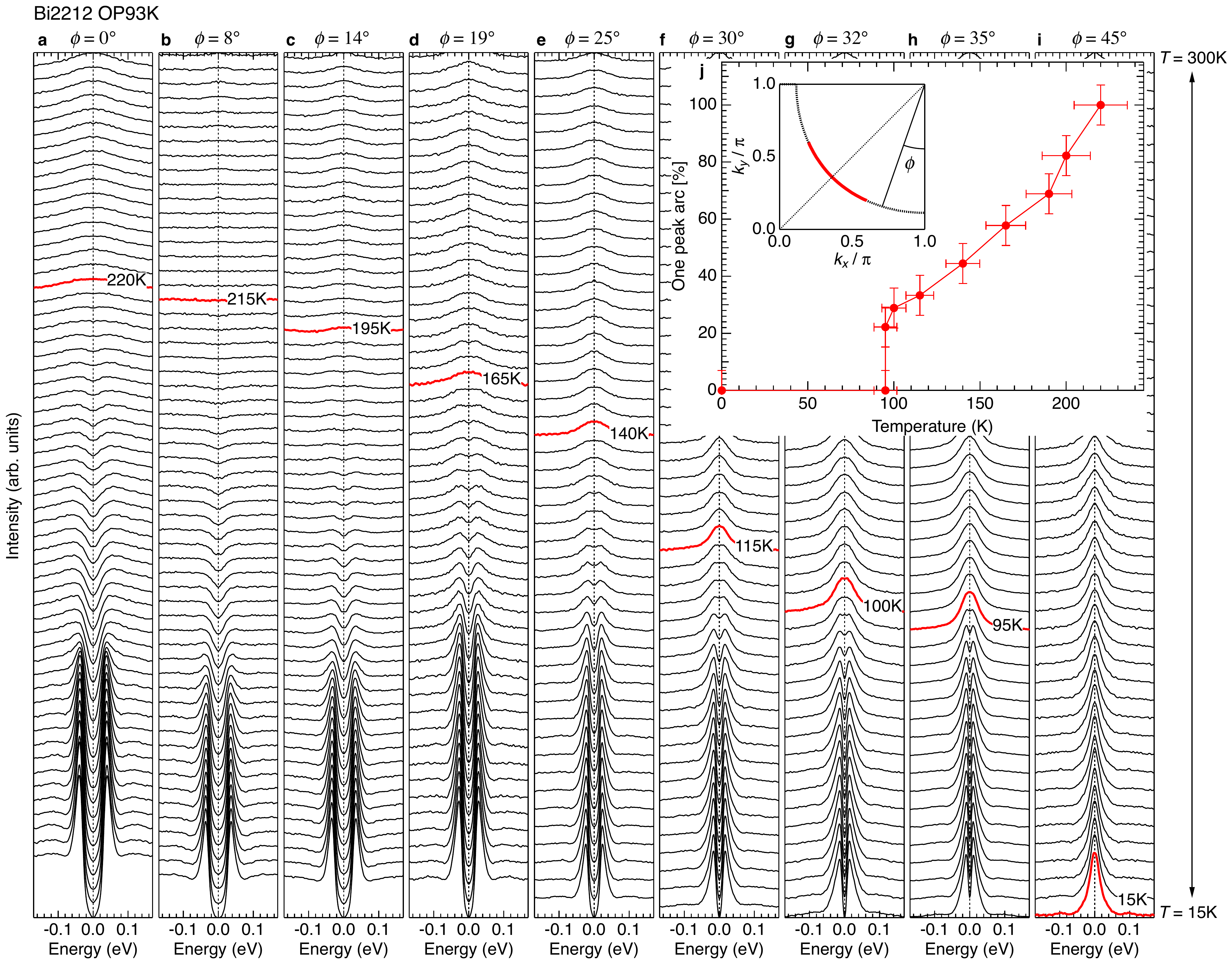}
\label{fig3}
\end{figure*}

\noindent {\bf Fig. S3.} 
The "artificial" Fermi arc of Bi2212 OP 93K obtained by tracking spectral peak. 
In each panel, the symmetrized EDCs measured  over a wide temperature range from below $T_{\rm c}$ to above $T^*$ are plotted.
From the left ({\bf a}) to the right ({\bf i}) panel, the measured $k_F$ point changes from the antinode to the node. 
The Fermi angle of the measured $k_F$ point [$\phi$: see inset of {\bf j}] is described on the top of each panel. 
The spectrum where the two peaks  merge to one peak at elevated temperatures is indicated with a red color. 
{\bf j}, The length of Fermi surface ("artificial" Fermi arc) where one peak structure is observed is plotted as a function of temperature.

\clearpage
\newpage

\begin{figure}
\includegraphics[height=8.5in]{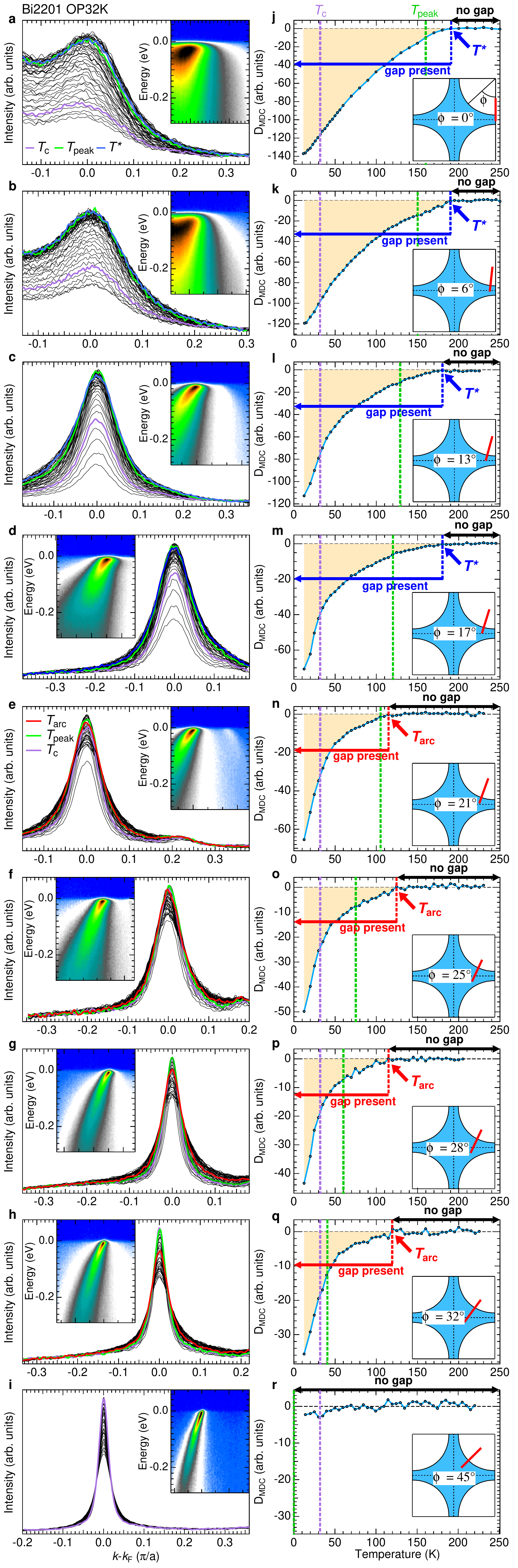}
\label{fig4}
\end{figure}
\clearpage
\newpage

\noindent {\bf Fig. S4.} 
Direction dependence of the partial density of state at $E_{\rm F}$, D$_{\rm MDC}$($E_{\rm F}$), for the optimally doped Bi2201 with $T_{\rm c}$=32K.
{\bf a-i}, MDCs at various temperatures from below $T_{\rm c}$ to above $T^*$. 
The top panel a corresponds to the antinodal cut. Toward the bottom  i, the momentum cut measured is shifted toward the node. 
 Each inset shows the band dispersion map, from which the MDC is extracted, at the lowest temperature (12K).
 {\bf j--r}, The temperature dependence of the partial density of state, D$_{\rm MDC}$($E_{\rm F}$), estimated from the area of MDC in a-i.
 The inset of each panel shows FS with a measured momentum cut, that is normal to FS. 
 The gap opening temperature ($T_{\rm DOS}$) estimated from the reduction of D$_{\rm MDC}$($E_{\rm F}$) and the temperature at which the symmetrized EDC has a peak at $E_{\rm F}$ are indicated with a blue and red dotted line, respectively. 

\clearpage
\newpage

\begin{figure}
\includegraphics[height=8.5in]{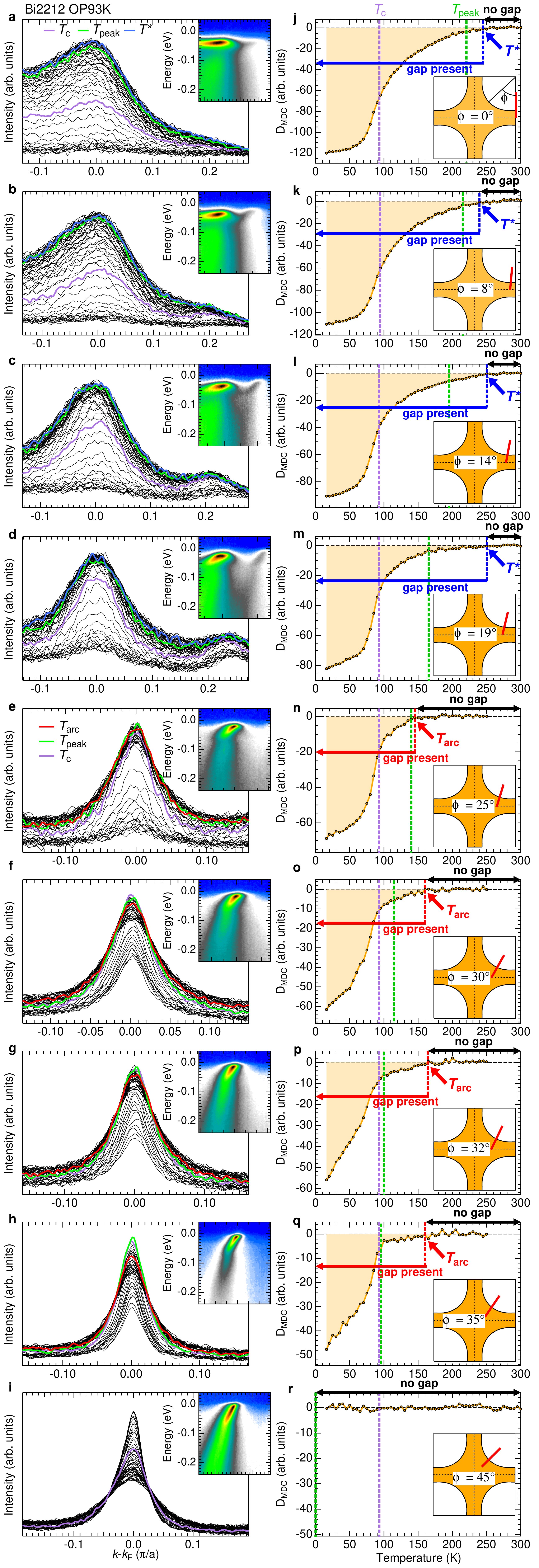}
\label{fig5}
\end{figure}
\clearpage
\newpage

\noindent {\bf Fig. S5.}
Direction dependence of the partial density of state at $E_{\rm F}$, D$_{\rm MDC}$($E_{\rm F}$), for the optimally doped Bi2212 with $T_{\rm c}$=93K.
{\bf a--i}, MDCs at various temperatures from below $T_{\rm c}$ to above $T^*$. 
The top panel a corresponds to the antinodal cut. Toward the bottom i, the momentum cut measured is shifted toward the node. 
Each inset shows the band dispersion map, from which the MDC is extracted, at the lowest temperature (12K).
 {\bf j--r}, The temperature dependence of the partial density of state, D$_{\rm MDC}$($E_{\rm F}$), estimated from the area of MDC in a--i.
 The inset of each panel shows the FS with the measured momentum cut, that is normal to FS.
 The gap opening temperature ($T_{\rm DOS}$) estimated from a reduction of D$_{\rm MDC}$($E_{\rm F}$) and the temperature at which the symmetrized EDC has a peak at $E_{\rm F}$ are  indicated with a blue and red dotted line, respectively.